# Nanoporous Metallic Network as a Largescale 3D Source of Second Harmonic Light

Racheli Ron[1], Omer Shavit[1], Hannah Aharon[1], Marcin Zielinski[2], Matan Galanty[1], and Adi Salomon[1]*

1. Department of Chemistry, Institute of Nanotechnology and Advanced Materials (BINA), Bar-Ilan University, Ramat-Gan 5290002, Israel.

2. Attolight AG, EPFL Innovation Park, Building D, 1015 Lausanne, Switzerland.

**Corresponding author**

*Dr. Adi Salomon. Email: adi.salomon@biu.ac.il.

**ABSTRACT**

Second harmonic generation (SHG) is forbidden from most bulk metals, because metals are characterized by a net zero electric-field in equilibrium. This limit breaks down when reaching nanoscale dimensions, as have been shown for metallic nano-particles and nano-cavities. Yet, nonlinear response from a *three-dimensional* (3D) macroscale *metallic* piece comprising sub-optical wavelength features remains a challenge for many years. Herein, we introduce a largescale nanoporous metallic network whose building-blocks are assembled into an effective nonlinear conductive material, with a considerable conversion efficiency in a wide range of optical wavelengths. The high nonlinear response results from the network structure having a large surface area on which the inversion symmetry is broken. In addition, because of the 3D structure, hot-spots can be formed also in deeper focal plans of the metallic network, and thus can give rise to coherent addition of the signal. The solid connectivity between the nanoscale building-blocks also plays a role, because it forms a robust network which may respond in a collective manner to form very intense hot-spots. Broadband responses of the metallic network are observed both by SHG and cathodoluminescence (CL). The large-scale dimension and generation of randomized hot-spots make this 3D metallic network a promising candidate for applications like photocatalysis, sensing, or in optical imaging as structured illumination microscopy.

1. Introduction

In a disordered network system, the building-blocks are randomly connected. This leads to a higher-level architecture, and the system may respond in a collective manner.[1,2] If large enough in size, the network becomes robust so that a local damage does not lead to a system failure.[3,4] Such robust disordered networks are of great importance in photonic devices, because they may respond in a cooperative manner to the applied electromagnetic (EM) field. In nature, corals are an example for such a robust disordered system, where light is diffusively scattered enabling symbiotic algae to capture a larger fraction of photons and optimize subaquatic photosynthetic energy production.[5]

Such 3D structures made of metals are absent in Nature, and their production, which comprises nanoscale features, remains challenging even with current technology.[6–11] The lack in techniques for preparing sufficiently pure and robust *metallic* networks has slowed down the optical investigation of these interesting disordered structures.[12] Nonetheless, the potential of such metallic networked systems both in photonics and in catalysis has been recognized.[7,11,13–16]

Three-dimensional disordered metallic nanostructures display a rich variety of optical phenomena.[2,7,11,17] Their random architecture with a broad nano-size and shape ranges of holes and particles results in EM field enhancement over a wide range of the optical spectrum. Moreover, their surface curvature and texture changes at the submicron scale, and therefore the EM field lines either converge or diverge, leading to refraction-index modulation at the submicron scale.[4,15] Typically, in such metallic networks there are regions with very high electron density, which thus exhibit strong local fields. These regions lead to excitation of localized plasmonic modes, and are referred as 'hot-spots'.[18] In addition, the connectivity between the building-blocks, such as nanoparticles, allows these networks to support propagating plasmonic modes.[19] As a result of surface plasmons (SPs) excitation, inherently weak nonlinear optical processes such as Raman scattering[16,20,21] and SHG[22–25] can be boosted by orders of magnitudes. Enhanced SHG from disordered metallic islands was already observed at 1981.[26] Nonetheless, thus far reported SHG emissions from such 2D disordered systems[27–37] as well as from 3D metallic system are rather low. Two main reasons are put forward to explain this observation; one is low number of hot-spots, which are generally disconnected and therefore their induced EM field does not overlap. The other one is the phase randomization of the emitted field that results from the incoherent hyper Rayleigh scattering (HRS) rather than the coherent SHG process.[23,38]

Recently, we have developed a technique to fabricate large-scale 3D disordered nanoporous metallic networks (see Figure 1).[9] The technique is based on physical vapor deposition (PVD) onto an electrostatic silica aerogel substrate (see Experimental section). These networks are made of chain-like ligaments with typical dimensions of about 100 nm, forming a continuous solid framework with multi-size and -shape pores. A key property of these disordered networks is that they combine a large number of two optically active elements: nano-pores and nanoparticles, both support plasmonic excitations over a broad optical frequency range. Nano-pores are sub-wavelength metallic cavities, and thus can trap the optical light.[39] The 3D architecture enables SH emission also from deeper plans, giving rise to enhanced signal due to constructive interference. Additionally, the connectivity between optical elements constructing the network should enable excitation of collective eigenmodes so to form intense hot-spots.[2]

Herein, we experimentally study SHG responses of a robust disordered 3D metallic network made of silver (Figure 1a). The SHG responses from such a network are relatively high and stable, in any given polarization state of the fundamental field and in several fundamental wavelengths (FWs). We observed emission of about 200 k counts per second (CPS) of blue photons, excited by 1.5 mW laser power with 0.5 NA objective in a reflection mode (see Figure 2a). Cathodoluminescence measurements support this observation, showing multiple emission hot-spots over a broad range of wavelengths.

## 2. RESULTS AND DISCUSSION

### 2.1. Characteristics of the SHG response from 3D silver network

Figures 1b-d display scanning electron microscopy (SEM) images of typical nanoporous silver networks. Three-dimensional entangled ligaments form a continuous percolating solid network. Sizes of the interconnected pores range from 5 nm to 500 nm, and thus permit permeability of both materials and energy. Morphologically, each ligament in the network is a string of metallic nano-size particles, around 100 nm thick, which are firmly connected by means of solid-fusion. This solid-fusion based connection between the metallic elements in the network assures its purity, and directly affects its optical performance. The variety in sizes and shapes of the optical elements forming the network, leads to a broadband spectral and high nonlinear responses, because both the fundamental and the emitted second harmonic fields can be potentially enhanced.

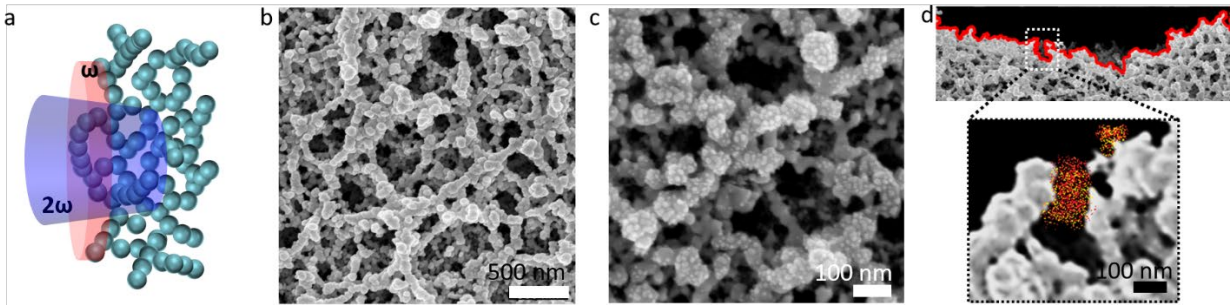

**Figure 1. The 3D metallic network**. **(a)** Illustration of a 3D disordered metallic network. When light impinges the nanoporous network surface, plasmonic modes can be excited at different frequencies due to broad size and shape ranges of both holes and particles. A concomitant EM field enhancement within the 3D silver network is expected both at the fundamental frequency ($\omega$) and at the SH frequency ($2\omega$). **(b-c)** SEM images of 3D silver network (top views). The average ligament size lies at about 100 nm, whereas cavities are multi-size (5-500 nm). **(c)** Networks are covered by <10 nm metallic tips, that strongly enhance light confinement at the metallic nanostructures constituting the 3D network. **(d)** Cross-sectional SEM image of the 3D silver network showing two different types of hot-spots as is illustrated for a nanoparticle and a gap. The scale bar is attributed for both the low- and high-magnification SEM images in **(d)**.

In order to study the nonlinear optical response of the network, samples were excited in reflection mode by a tunable Ti:Sapphire laser (690–1080 nm, 100 fs pulse duration, 80 MHz repetition rate) using ×50/0.5NA air objective at normal angle (see Experimental section). A schematic illustration of the SHG set-up is shown in Figure 2a. In short, samples were mounted on a piezoelectric stage, and were scanned along both x and y axes. The laser power was set at 1.5 mW at the laser exit for most of the experiments, yet, we measured also with higher laser power. The SHG responses along two perpendicular polarization directions were measured by directing the reflected light via a polarizing beam splitter onto two avalanche photodiodes (APDs). Reliability of our SHG systems was confirmed both by measuring a quadratic relation between input fundamental laser power and output SHG responses (Figure 2b), and by spectral measurements as is further detailed in the following.

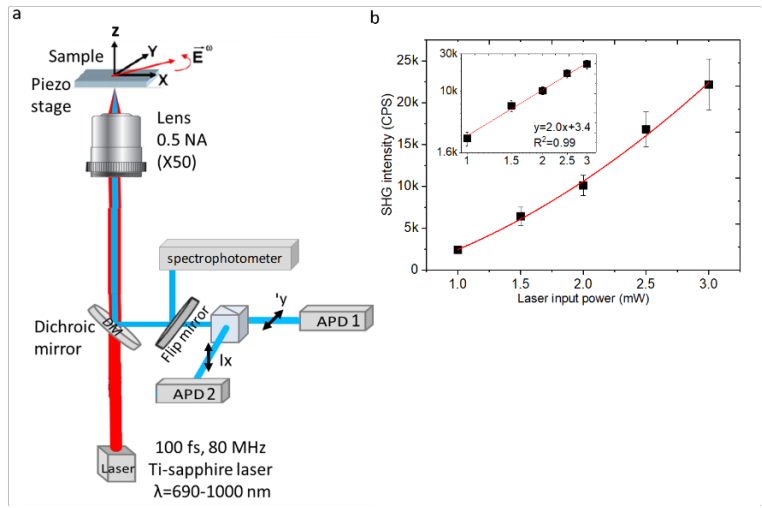

**Figure 2. Illustration of our SHG experimental set-up. (a)** A focused beam from a Ti-sapphire laser of 100 fs pulse duration and 80 MHz repetition rate is directed to the sample using 0.5 NA and ×50 magnification objective, and the SH signal is collected in reflection mode by the same objective. The reflected SH is directed to a beam splitter and collected by two perpendicular APDs. For spectroscopic measurements, the beam can be directed to the spectrometer using a flip mirror. **(b)** Quadratic relation between input fundamental laser powers and output SHG responses is resulted, as expected for such nonlinear process. The inset shows the log-log representation of SHG versus excitation power relation. Slope of about 2 between was achieved with 99% linear fitting.

Characteristic SHG emission scan from a silver network is shown in Figure 3. Clearly, an intense SHG emission is observed from certain confined areas, the so-called 'hot-spots'. High-resolution SEM images revealed that the network ligaments are covered by <10 nm tip-shaped nanoparticles as is clearly seen in Figure 1c. This thorny texture contributes significantly to enhancing EM field densities at these sharp metallic tips. The hot-spots are spatially localized into sub-micron scale areas, and their apparent size is diffraction-limited by our set-up. Magnitudes of the emitted SHG signal from the silver network, without considering losses along the optical path, are in the order of 20,000 CPS. For comparison, the SHG signal from a compact continuous silver film exhibits less than 500 CPS from both APDs under the same FW power settings (1.5 mW, 0.5 NA objective), lying close to the noise level of the APDs, see SI Section S*A*. That is, these metallic networks generate an enhanced SHG emission, which is at least 200 times stronger comparing to a thin silver film (see detailed information in SI Section S*A*).

The fluctuating SHG emission behavior for a *single FW* of 940 nm is shown in Figures 3c-f, which compare the average SHG signal at different 0.5×0.5 µm$^2$ network areas under 1.5 mW power. The mean SHG emission intensity values present a small variation for certain hot-spots (compare A, B, C spots in Figure 3d), and variation of an order of magnitude within other locations (compare, for example, B and F spots in Figure 3d).

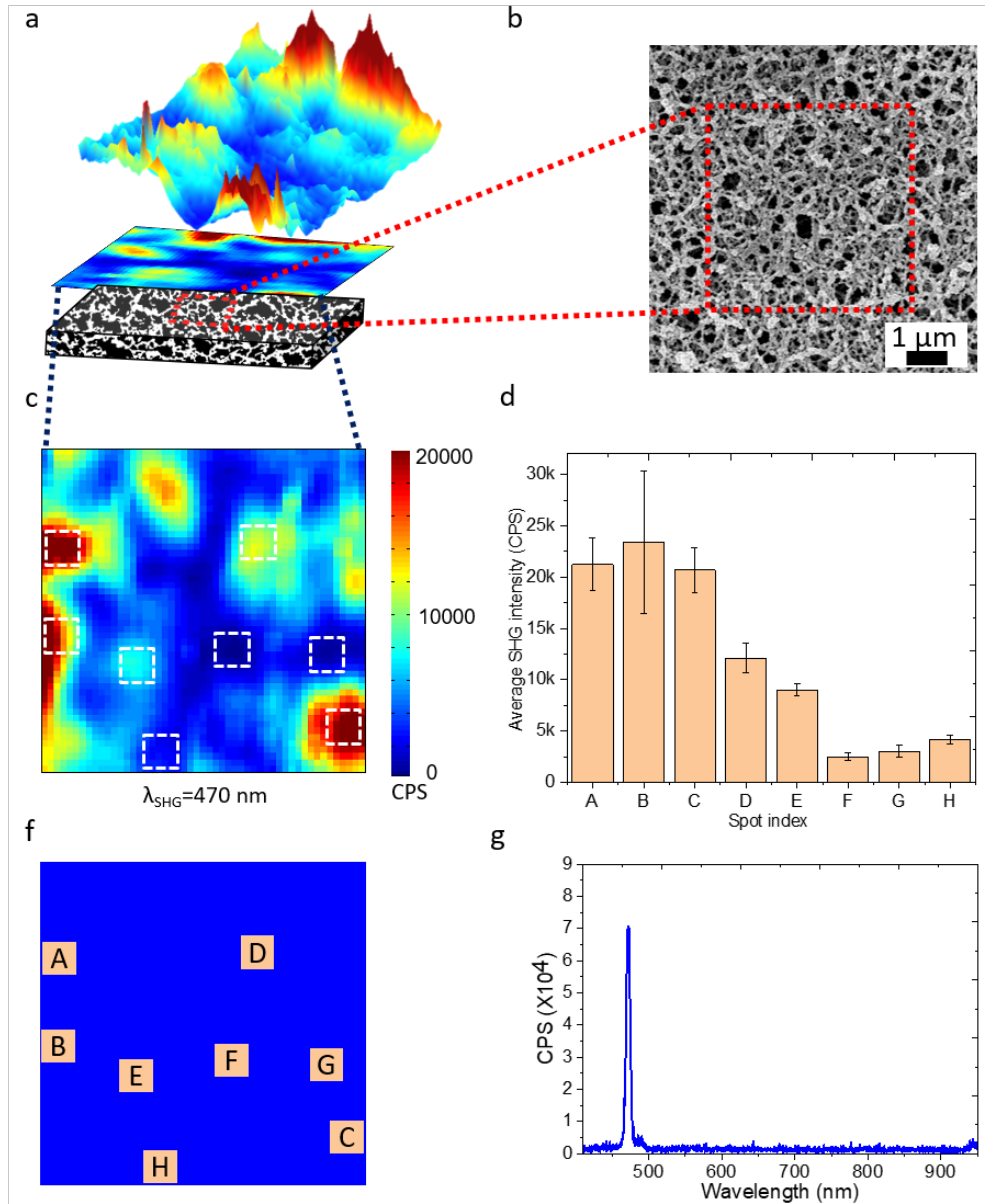

**Figure 3. SHG response of the metallic network. (a-c)** SHG response of the nanoporous silver network over 5 × 5 μm² scan area under 940 nm laser illumination and 1.5 mW input power, as is indicated in the SEM image in **(b)**. **(d)** Average SHG emission from various 0.5 × 0.5 μm² network locations, marked by white dashed squares in **(c),** and are also illustrated in **(f). (g)** SHG emission spectrum under incident fundamental beam of 940 nm indicating that processes other than SHG are negligible.

Considering the inversion symmetry of the silver lattice, only negligible SHG response is expected due to symmetry breaking at the surface. However, our reported silver networks combine two optically active elements: metallic nanoparticles and nano-cavities, the two support localized SP excitations over the whole optical range- both at the fundamental and at SH frequencies. The 3D organization of the network elements as well as the excessive surface area, are reasons for such an efficient SHG responses. In addition, the connectivity between the optically active building-blocks may enable collective responses throughout the network. Inasmuch, the collective optical behavior of the network is currently under study. As well, we note that in such disordered system a diffused scattering may be observed leading to high SHG intensity towards the backscattering direction (see SI Section S*B*).[40,41]

Typical spectral characteristics of the emitted SHG signal under measurement conditions of 1.5 mW and 940 nm FW are shown in Figure 3g. A sharp SHG peak at 470 nm is observed with minute residual signal of the fundamental field pointing out the reliability of our optical system. No luminescence or other multi-photon processes are observed in this case, indicating that the collected photons resulted solely from SHG. The expected quadratic intensity dependence of SHG is shown in Figure 2b. For comparison, we measured the SHG emission spectrum from a network upon higher input laser power of 5 mW and 940 nm incident fundamental beam (see Figure S3 in SI section *SC*). An intense SHG related peak at 470 nm is observed together with other minor nonlinear processes above and below 470 nm. In addition, stability measurement of the emitted SHG signal from 3D silver network versus time is shown in Figure S4 excluding photo-damages.

The SHG conversion efficiency from the silver networks was evaluated to be $5.6 \times 10^{-11}$, taking into account losses along the detection path in our experimental set-up and using quite low input power of the fundamental beam. See SI Section *SD* for the detailed calculations. This resulted value is relatively high compared to reported SHG conversion efficiencies in other works *at this* *optical region*. Well defined and highly engineered metallic nanostructures are reported to be efficient at the IR region[42–47], a region in which the presented 3D network has not been studied yet.

To emphasize the structural uniqueness of these metallic networks, we measured the SH response from a denser silver network, which is mostly lack of percolating pores. A typical SHG emission from a denser silver network is displayed in Figure 4. Of note, localized sub-micrometer scale hot-spots are neither detected from continuous silver films nor from denser silver networks. Clearly, the network porosity and density affect its optical performance. The SHG response from a denser silver network is relatively uniform over the scanned area with minor hot-spots, and introduces a lower average signal which is one order of magnitude smaller compared to the SHG signal from the studied nanoporous networks.

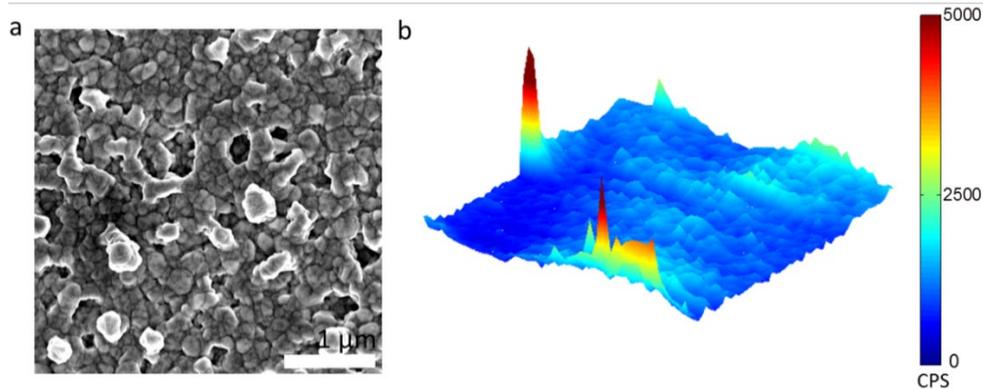

**Figure 4. Effect of the network density on SHG responses**. **(a)** SEM image of a dense silver network. **(b)** SHG response of a typical dense silver network ($5 \times 5$ μm$^2$). The SHG signal is relatively uniform over the scanned area with minor hot-spots and a lower average signal of about 2500 CPS. Measurements were carried out under identical conditions of 940 nm fundamental wavelength and 1.5 mW power. Two-dimensional presentation of the SHG scan is shown in SI section *SE*.

The network nanostructure is accompanied by giant refraction index fluctuations at the sub-micrometer scale. This refraction index inhomogeneity induces a prominent EM localization, which is mediated by plasmons excitation.[4,15] Concentration of the EM energy at sharp nanometer size tips form thereby hot-spots of peak field intensities. These intense EM peaks function as local nonlinear sources for SHG. Nevertheless, we believe that the network structure permits connectivity and cooperativity between the metallic sources generating larger hot-spots.

The 3D architecture provides the nanoporous network with a distinct advantage over other conventional plasmonic structures[24,46–48], in regards with the available depth-of-field for SHG emission. Figure 5 shows the SH intensity from different focal planes in the network which were generated from a constant network location. Controlled movements along the microscope z-axis was done using a motorized focus drive. It is observed that an average intensity of about 20 k CPS (without considering losses) is preserved while varying the focal position at the specimen over 6 µm range. Not all network locations are identical in their response to a focal change. There are network locations in which the focal distance can be varied along 60 µm while the signal intensity is preserved, yet with 20% deviation from the average intensity. For comparison, the intensity of the SHG signal drastically decrease upon changing the focal plane by 3 µm for a 2D plasmonic hole-array structure milled in a silver film (not shown).

The ability of the network to emit SH light from multiple plans demonstrates the fact that the network is a 3D source for SHG. Implications of this 3D network are multiple for functional purposes, one example is using such network as an efficient platform for 3D imaging and bio-sensing.

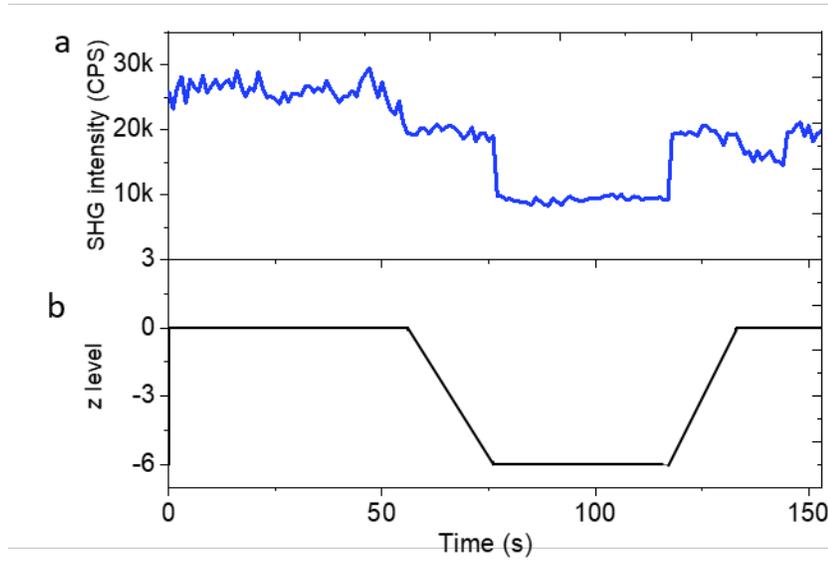

**Figure 5. SHG intensity as function of the focal plane in the 3D silver network. (a)** SHG emission intensity as function of the focal plane in the network, which is represented versus time by the graph in **(b)**. The "0" z-level in **(b)** notes the level of the initial focal plane (surface), and the rest of the distances are noted in accordance. The SHG response was recorded for 940 nm illumination, and 2 mW input power. The signal was recorded using one polarization channel of the emitted signal while neglecting the rest.

**2.2. Broadband SHG response and spatial localization**

The multiple sizes and shapes of the metallic nanoparticles and nano-holes result in broadband excitation of plasmonic modes in both visible and near-IR regions of the spectrum.[2,49] Therefore, we expect the silver network to enhance the SHG response at multiple FWs. We shall note that plasmonic modes both at the fundamental and at the SH frequencies can be excited, due to broadband responses of the metallic network.[9,10]

To further investigate the nonlinear response of the silver networks, we performed wavelength-dependent SHG measurements. Figure 6 shows SHG emissions from the *same* 5 × 5 μm$^2$ silver network upon illumination with three different fundamental wavelengths of 900, 940 and 980 nm. All the other parameters were kept identical, and these wavelengths have been chosen so that the response of our set-up (APDs, filters) is very similar. Contour-lines were added at identical locations over the three different SHG scans, to facilitate comparison between the wavelength-dependent SHG emissions. An enhanced and localized SHG emissions are detected for all three wavelengths. Yet, the hot-spots appear at different areas of the network are dependent on the fundamental excitation wavelength. For example, the *bottom-right* hot-spot (yellow square) has its strongest intensity appeared for 940 nm fundamental illumination compared to the other FWs. Whereas, a lower SHG response from this bottom-right network location was observed for 900 nm, and a negligible intensity for 980 nm. We observed intensity differences of about half order of magnitude upon illumination of the network by the three different FWs. This "on/off" behavior, coming into expression by a fluctuating SHG emission as function of the exciting FW, might be useful for structured illumination microscopy (SIM), for example.[50] This is because the hot-spots can be alternatively activated as function of the FW (and polarization). Additionally, since the average size of the building-blocks can be controlled by the physical parameters of the fabrication,[9,10] some tuning of the SHG responses for a specific wavelength might be attained as well.

In order to quantitatively evaluate the dependence of SHG response on the exciting wavelength, average SHG intensities for each FW were calculated (Figure 6d). Considering the whole 5 × 5 µm² scanned network area (Figures 6a-c), the highest average SHG emission was generated under 940 nm illumination. More interesting, we calculated average SHG emissions from two different 1 × 1 µm² network areas; one is of rather *low* emission noted by central white squares in Figures 6a-c, and two other locations of *fluctuating* SHG emission as function of the illumination wavelength, noted by bottom-right and upper-right yellow squares. The central spots generate a relatively lower SHG emission under the three different FWs comparing to the two other mentioned hot-spots. However, their average SHG intensity is on the order of 2000-4000 CPS (without considering losses), which is higher emission level compared to previous works measured SHG from silver nanoparticles or nano-holes[24,46].

The bottom-right yellow squares introduce an "on/off" behavior of the SHG responses as function of FWs. That is, the average SHG intensity is ~16,000 CPS for 940 nm FW, which is about ×2 and ×4 times higher than for 900 and 980 nm FWs, respectively. These results demonstrate the broadband nonlinear response of a single 3D nanoporous silver network.

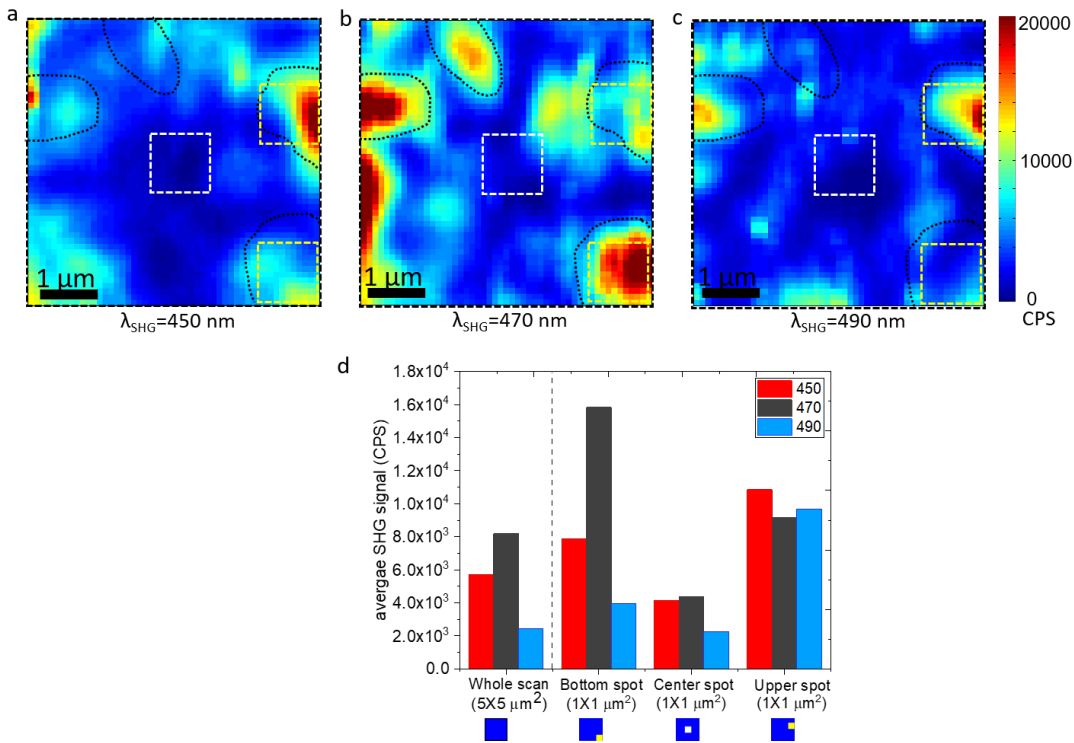

**Figure 6. Wideband visible SHG response of 3D silver network**. Scans of SHG responses emanating from the same 3D silver network (5×5 μm$^2$) under illumination with three different FWs of **(a)** 900, **(b)** 940 and **(c)** 980 nm. Although from the same network area, hot-spots are observed at different locations. The added contour lines are aimed at facilitating comparison of the nonlinear responses of specific network locations under the different FWs. **(d)** Mean SHG responses from the same silver network under three different FWs. The histogram presents mean SHG intensity from the whole scan (left), and from three different network areas, for each of the examined fundamental wavelengths of **(b)** 900 nm **(c)** 940 nm, and **(d)** 980 nm. Locations are indicated by colored squares: *white* 1×1 μm$^2$ center spot, *yellow* 1×1 μm$^2$ bottom-right and upper-right spots (see sketches below).

## 2.3. Linear optical behavior

The broad optical responses of the nanoporous silver networks are demonstrated by linear optical[9,10] and CL measurements. The wide optical response of the metallic network is coming into expression by spots of multipole colors observed by the optical reflection micrograph in Figure 7a. Yet, the overall resulting color of the large-scale silver network is gold-like yellowish (see the inset in Figure 7a).

In order to better resolve the linear optical properties, we used CL which permit nanoscale resolution. In this technique, the specimen is irradiated by an electron beam, and the ejected visible photons are collected. Figure 7b displays CL image of the network whose SEM image presents in Figure 7c. The CL image comprises 3 stacked monochromatic CL images of the network at $\lambda=460\pm5$ nm, $540\pm5$ nm, and $740\pm30$ nm. It can be observed that CL emission was collected from all the scanned network area, and that different network locations give rise to different colors. More specifically, CL spectra from three nearby network locations (Figure 7c) are presented in Figure 7d, revealing plasmonic eigenmodes at various frequencies. We attribute the observed spectra to the multimode SP excitations, both from the particles and the pores. The excitation of large number of SP modes increases the density of photonic modes at the interface leading to the observed enhanced nonlinear responses at several FWs.

The broadband linear behavior of disordered metallic networks is known to extend into the IR[2] and as evidenced from linear optical measurements,[9] such property is meaningful for boosting SHG because it supports mode matching both at the fundamental excitation wavelength and second harmonic wavelength.[45] Hence, the results from both the CL and linear optical analyses point that silver in the form of a nanoporous 3D network is a spectrally dense medium, with confined EM fields locations of energy-modes forming hot-spots. In particular, the disordered network is a multi-resonant architecture and can support intense SHG at several optical wavelengths.

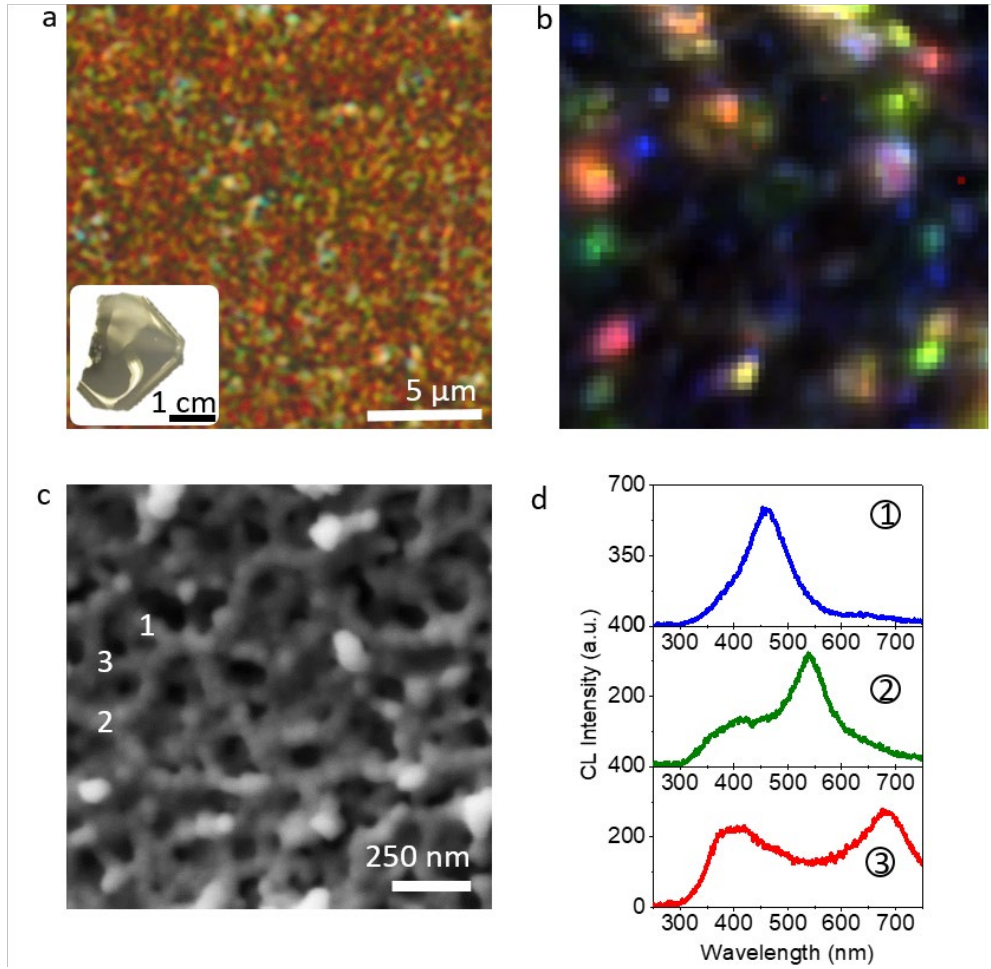

**Figure 7. Linear optical behavior of a nanoporous silver network**. **(a)** Bright field optical micrograph in reflection of 3D silver network. The inset is an optical-camera photograph of a typical silver network, showing its overall large-scale dimension and gold-like yellowish color. **(b-d)** Cathodoluminescence emission from 3D silver network. **(b)** CL image of the silver network which is shown by the SEM image in **(c)**. The CL image in **(b)** is made from stacking three monochromatic CL images at λ=460±5 nm, λ=540±5 nm, and λ=740±30nm. **(d)** Point CL spectra which were extracted from different spatial locations at the silver network, marked by numbers on the SEM image of the analyzed network in **(c)**.

### 3. CONCLUSIONS

Large-scale nanoporous silver networks show a considerable high nonlinear response, both compared to flat silver surface and compared to conventional well-defined and highly-engineered plasmonic nanostructures. Furthermore, a confinement of the SHG field into diffraction-limited hot-spots is observed. The network, which is an actual 3D structure, can function as a macroscopic 3D "carpet" of hot-spots, and thus can lead to enhancement of photochemical reactions of molecules located in proximity, or to enhanced detection of molecules. In the context of strong-coupling and energy-transfer processes between plasmonic modes and molecular transition states,[51–53] the metallic network offer a new platform because it can induce long-range interaction between the molecules themselves. Thus, the network plasmonic modes may induce coupling between near and remote molecules and can serve as a 'pool' which shape the energy transfer between the molecules themselves.[54,55]

## 4. EXPERIMENTAL SECTION

### 4.1. Silver network sample preparation

Samples of 3D nanoporous silver networks were prepared by sputtering on a silica aerogel substrate, as is detailed elsewhere.[9] In brief, samples preparation comprises two steps: (1) aerogel substrate synthesis, and (2) PVD. Silica aerogels were synthesized by a one-step base-catalyzed sol-gel process followed by $CO_2$ supercritical drying (K850, Quorum). Metals were sputtered (682, Gatan) from pure targets (99.99%, kurt Lesker).

### 4.2. SHG measurements

A schematic illustration of the SHG set-up is shown in Figure 2a. The SHG set-up includes a tunable Ti:sapphire laser (Mai-Tai HP, Spectra Physics, 690-1080 nm) with 100 fs pulse duration and 80 MHz repetition rate. Samples were mounted on a piezoelectric stage with a closed-loop feedback (Piezosystem Jena), and were scanned with 100 nm step-size along both x and y axes.

Controlled movements along the microscope z-axis was done using a motorized focus drive (Marzhauser Wetzlar).

The laser power was controlled by a Glan-Taylor polarizer, and was set at 1.5 mW. The incident FW was focused on the sample surface by an inverted microscope using a ×50/0.5NA air objective (Olympus) at normal angle. The nominal spot size is 1.1 μm. Detailed calculation and the specific spot size value as function of wavelength are presented in SI Section SD. The SHG signal was collected in reflection mode through the same objective. A dichroic mirror (Chroma) was used to block the reflected fundamental beam, and appropriate band-pass filters were used to further isolate the SH radiation. The reflected light was directed via a polarizing beam splitter onto two APDs (Perkin-Elmer) to measure the SHG intensity along two perpendicular polarization directions. The nominal dark noise level of the APDs is 200 CPS. To measure the spectrum, the emitted SHG signal was directed into the entrance slit of a spectrograph (303i, Shamrock) equipped with an electron multiplied charge-coupled device camera (Andor Newton).

### 4.3. Linear optical and CL measurements

Optical imaging was carried out using an inverted light microscope (BX51, Olympus) in reflection mode employing a ×100/0.90NA air objective (Olympus).

CL measurements were performed by an Attolight Rosa 4634 CL microscope (Attolight AG, Switzerland), which tightly integrates a high numerical aperture 0.72 NA achromatic reflective lens within the objective lens of a field-emission-gun scanning electron microscope (FEG-SEM). The focal plane of the light lens matches the FEG-SEM optimum working distance. CL was spectrally resolved using a Czerny-Turner spectrometer (320 mm focal length, 150 grooves per mm grating) and measured with a UV–visible CCD camera. Acceleration voltage of 10 kV, and electron beam emission current of 20 nA were applied.

**SUPPORTING INFORMATION**

Supporting information is available.

**ACKNOWLEDGEMENTS**

We thank prof. M. Oheim for carefully reading the manuscript.

**FUNDING**

This work was supported by the Israeli Prime Minister office via the INREP project, and by the Energy and Water Resources Ministry of Israel grant number 216-11-016.

**KEYWORDS**

Second harmonic generation, nanoporous metals, disordered robust optical system, hot-spots, plasmonics, cathodoluminescence.

# Supporting Information

# Nanoporous Metallic Network as a Largescale 3D Source of Second Harmonic Light

Racheli Ron[1], Omer Shavit[1], Hannah Aharon[1], Marcin Zielinski[2], Matan Galanty[1], and Adi Salomon[1]*

1. Department of Chemistry, Institute of Nanotechnology and Advanced Materials (BINA), Bar-Ilan University, Ramat-Gan 5290002, Israel.
2. Attolight AG, EPFL Innovation Park, Building D, 1015 Lausanne, Switzerland.

## SA. Comparison between SHG responses emanating from a silver film versus a 3D silver network

SHG signals at orthogonal polarizations from a silver film and a silver network are presented in Figure S1.

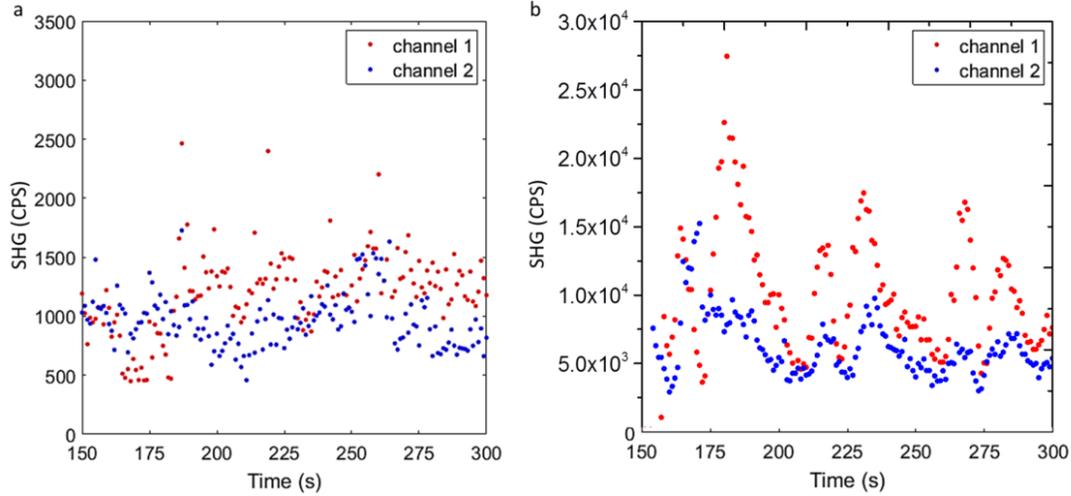

**Figure S1.** SHG response from **(a)** thin silver film and **(b)** a 3Dsilver network. SHG signal was collected by two APDs (channels 1 and 2) placed at orthogonal polarizations under fundamental wavelength illumination of 940 nm. Power of 10 mW was applied for the silver film measurement, comparing to silver networks measurements, in order to be able to detect a clear signal. Note the different scales in **(a)** and **(b)**, which are aimed at facilitating observation.

Comparing SHG response of a silver network versus that of a thin compact silver film upon identical measurement conditions, an enhancement of ×200 is obtained, see Table S1.

**Table S1. SHG responses from a silver film and a silver network,** *without taking losses into consideration.*

|  | Input power (mW) | Measured SHG signal* (CPS) |
|---|---|---|
| **Silver film** | 1.5 | 100** |
|  | 10 | 2,000 |
| **Silver network** | 1.5 | 20,000 |

*Sum of two orthogonal polarization channels.
**Under 1.5 mW illumination power, the compact thin silver film generates signal which is under the APD`s noise level. The noise level of each APD is 200 CPS, therefore we cannot draw conclusions from such measurements. Hypothetically, in case that the APDs would not have a dark noise, one would have expected to measure about 50 CPS (as calculated from the measurements of 10 mW, according to $I_{(1)2\omega} = \frac{(I_{(2)\omega})^2}{f^2}$, while f is the ratio between higher to low input powers.

**SB. Directional reflection from disordered metallic nanostructures**

Pronounced multiple scattering occurs in a disordered media. Reflection through these nanoporous metallic networks does not necessarily decay exponentially with thickness.[1,2] With respect to scattering, both the mean pore size and the mean grain size are expected to play important roles, because they determine how photons migrate through the network. The pore size will determine the scattering mean free path whereas the particle size will impact the scattering anisotropy, that is, the distribution of scattering angles following each scattering event. If the pores are large, the free path between two scattering events might be large enough to permit faint scattering, leading to closed loop photon trajectories and *enhanced reflection*. In this case, the output intensity is relatively high, because the probability of photons to exit the material with a trajectory similar to the original one, after multiple scattering events, is relatively high.

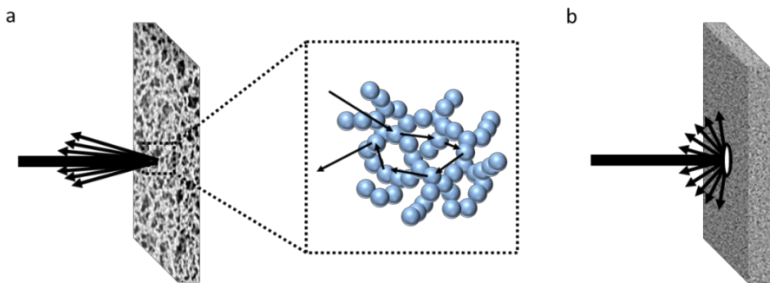

**Figure S2. Light propagation in disordered structures.** When pore sizes are large enough closed loop photon migration might arise. **(a)** Light scattering through metallic networks can be directional towards the backscattering direction as opposed to **(b)** widely scattered.

**SC. Spectral characterization and network stability**

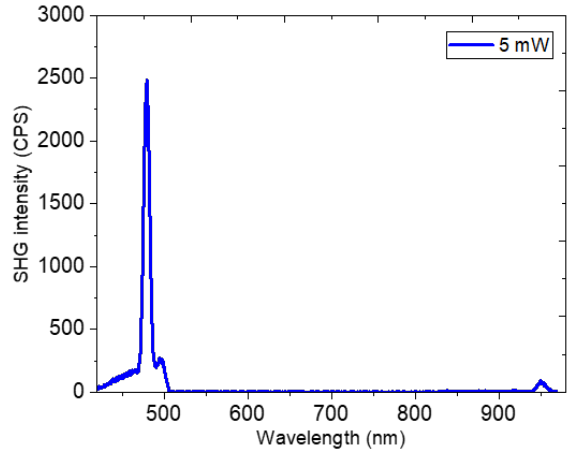

**Figure S3. Spectral characterization of SHG emission upon higher input laser power.** SHG emission spectrum under 5 mW power and 940 nm incident fundamental beam. An intense peak at 470 nm, the SHG related wavelength, is observed together with other nonlinear processes such as luminescence (bellow and above 470 nm). It worth mentioned that the metallic network is stable under this laser power.

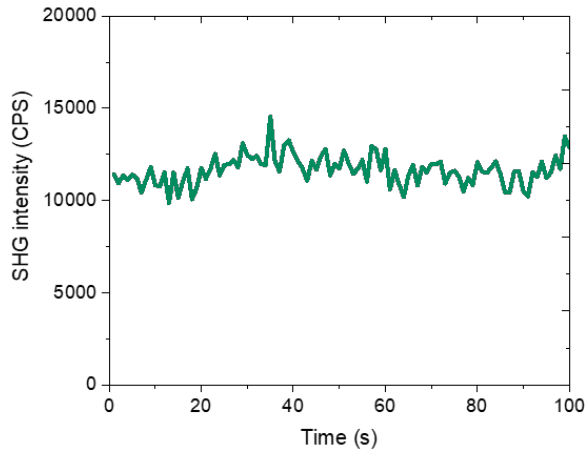

**Figure S4. Stability of the emitted SHG signal from 3D silver network.** SHG emission intensity from a 3D silver network versus time upon illumination with 2 mW input laser power at 940 nm wavelength. The average SHG intensity along 100 s is about 11,500 CPS.

## SD. Losses consideration in our set-up

Input SHG experimental parameters and laser power calculations are presented in Table S2.

**Table S2. Input experimental parameters.**

| Parameter | Fundamental wavelength (nm) | Laser repetition rate (MHz) | Laser pulse length (fs) | Objective numerical aperture | Laser power (average) (mW) | Laser peak power (W) |
|---|---|---|---|---|---|---|
| Symbol/ Equation | $\lambda$ | $\nu$ | $\tau$ | NA | $<P_{FW}>$ | $\widehat{P}_{FW} = \frac{<P_{FW}>}{\tau \nu}$ |
| Value | 940 | 80 | 100 | 0.5 | 1.5 | 187.5 |

Average intensity of the irradiating laser beam is $<I_{FW}> = \frac{<P_{FW}>}{A_{obj(\omega)}} = 3.63 \times 10^8$ (W/m²). $A_{obj}$ denotes the spot area of the objective lens. The exciting spot diameter is $\delta(\omega) = \frac{1.22\lambda}{NA} = 2.29$ (μm) for $\lambda_{FW}$=940 nm. The exciting spot area at the focus of the laser is $A_{obj} = \pi(r_\omega)^2 = 4.13 \times 10^{-12}$ (m²), $r_\omega$ is the radius of the illuminated area under FW illumination. Peak intensity of the irradiated laser is $\hat{I}_{FW} = \frac{\widehat{P}_{FW}}{A_{obj(\omega)}} = 4.54 \times 10^{13}$ (W/m²), with regards to the laser peak power of 625 W and $A_{obj}$ at 940 nm.

We evaluated our system transmittance to be able to compare our SHG enhancement with others. To evaluate efficiency of SHG emission from the 3D silver networks, we considered the losses along the optical path length in our experimental set-up. First, the total transmittance of our experimental set-up at 470 nm is evaluated; The transmittance of the objective is ~30% with respect to a collection angle of ~1.5 steradian. Then, only 70% of the incoming light is transmitted through the optical components including lenses, four filters, dichroic mirror, and polarizing beam splitter. The efficiency of the APD is 60% at 470 nm. About 80% of the light is coupled to the fiber. Consequently, total transmittance efficiency of our system at 470 nm is ~ 10%.

Considering the actual total transmittance of our experimental system, the average SHG emission from the silver networks is higher than ***200 kCPS*** at 0.5×0.5 μm² hot-spot area. Output SHG power calculations and values are detailed in Table S3. Therefore, we arrive at a peak nonlinear coefficient of $\gamma_{SHG} = 3.01 \times 10^{-13}$ (W$^{-1}$), and conversion efficiency of $\eta_{SHG} = 5.64 \times 10^{-11}$ from the silver networks (Table S4).

**Table S3. Output SHG power calculations considering losses\*.**

| Parameter | SHG emitted photons (CPS) | SHG photon energy (J) | SHG Power (W) | SHG Peak power (W) |
|---|---|---|---|---|
| Symbol/ Equation | $\frac{N}{\Delta t}$ | $E_{ph,SHG} = \frac{hc}{\lambda_{SHG}}$ | $<P_{SHG}> = \frac{N}{\Delta t} \cdot E_{ph,SHG}$ | $\widehat{P}_{SHG} = \frac{<P_{SHG}>}{\tau \nu}$ |
| Network | 200,000 | 4.23 ×10$^{-19}$ | 8.46×10$^{-14}$ | 1.06 ×10$^{-8}$ |

\* For 470 nm, at hot-spot.

$E_{ph,SHG}$ denotes energy of the emitted SHG photon at $\lambda_{SHG} = 470\ nm$, h=6.63×10$^{-34}$ J•s is Planck constant, and c= 3×10$^8$ m/s is the speed of light in vacuum.

**Table S4. SHG efficiency calculations considering losses.**

| Parameter | Conversion efficiency | Peak Nonlinear Coefficient (W$^{-1}$) |
|---|---|---|
| Equation | $\eta_{SHG} = \frac{<P_{SHG}>}{<P_{FW}>}$ | $\gamma_{SHG} = \frac{\widehat{P}_{SHG}}{(\widehat{P}_{FW})^2}$ |
| Value | 5.64 ×10$^{-11}$ | 3.01 ×10$^{-13}$ |

[1]For 470 nm. Signal at a hot-spot.

**SE. Effect of network density on SHG response**

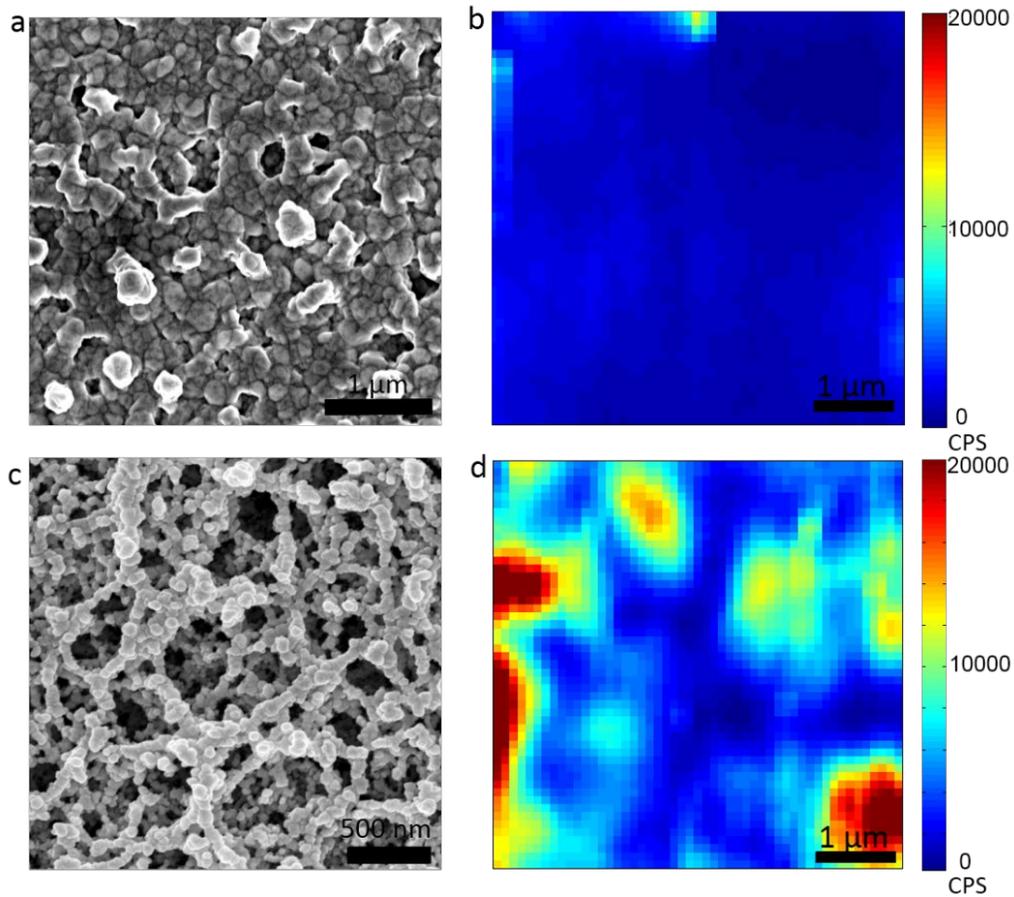

**Figure S5**. SHG response from a dense silver network. **(a)** SEM image and **(b)** SHG scan of a dense silver network (5×5 μm² area). To facilitate comparison, our typical 3D silver network is shown in **(c)** with its typical SHG response in **(d)**. SHG signal from the dense network is relatively uniform over the scanned area with minor hot-spots.

**Supporting References**